%% file: main.tex
\documentclass[sigconf]{acmart}
\usepackage[T1]{fontenc}
\usepackage{booktabs} 
\usepackage{multirow}
\usepackage[utf8]{inputenc}
\usepackage{url}
\usepackage{graphicx}
\usepackage{color}
\usepackage{tablefootnote}
\usepackage{footnote}
\usepackage{balance}
\usepackage{tabularx}
\usepackage{textcomp}
\usepackage{caption}
\usepackage{subfigure}
\usepackage{footnote}
\usepackage{enumitem}

\usepackage{multirow}
\usepackage{amsmath}
\definecolor{Gray}{gray}{0.9}
\usepackage{colortbl}
\usepackage{balance}

\AtBeginDocument{%
  \providecommand\BibTeX{{%
    \normalfont B\kern-0.5em{\scshape i\kern-0.25em b}\kern-0.8em\TeX}}}

\copyrightyear{} 
\acmYear{} 
\setcopyright{none}
\acmConference{*This is a preprint version of an accepted paper on WebMedia. Please}{}{consider to cite the conference version instead of this one.}
\acmPrice{}  
\acmDOI{}
\acmISBN{bla}

\begin{document}

\title{Helping Fact-Checkers Identify Fake News Stories Shared through Images on WhatsApp*}

\author{Julio C. S. Reis} 
\orcid{0000-0003-0563-0434}
\affiliation{%
  \institution{Universidade Federal de Viçosa}
  \city{Viçosa}
\country{Brazil}
}
\email{jreis@ufv.br}

\author{Philipe Melo}
\orcid{0000-0001-9830-1896}
\affiliation{%
  \institution{Universidade Federal de Viçosa}
  \city{Florestal}
\country{Brazil}
}
\email{philipe.freitas@ufv.br}

\author{Fabiano Belém}
\orcid{0000-0003-1076-2052}
\affiliation{%
  \institution{Universidade Federal de Minas Gerais}
  \city{Belo Horizonte}
\country{Brazil}
}
\email{fmuniz@dcc.ufmg.br}

\author{Fabricio Murai}
\orcid{0000-0003-4487-6381}
\affiliation{%
  \institution{Worcester Polytechnic Institute}
  \city{Worcester}
\country{USA}
}
\email{fmurai@wpi.edu}

\author{Jussara M. Almeida}
\orcid{0000-0001-9142-2919}
\affiliation{%
  \institution{Universidade Federal de Minas Gerais}
  \city{Belo Horizonte}
\country{Brazil}
}
\email{jussara@dcc.ufmg.br}

\author{Fabricio Benevenuto}
\orcid{0000-0001-6875-6259}
\affiliation{%
  \institution{Universidade Federal de Minas Gerais}
  \city{Belo Horizonte}
\country{Brazil}
}
\email{fabricio@dcc.ufmg.br}

\renewcommand{\shortauthors}{Reis, et al.}

\begin{abstract}
\input{src/00-abstract}
\end{abstract}



\maketitle

\input{src/01-introduction}

\input{src/02-background}
\input{src/03-datasets}
\input{src/04-features}
\input{src/05-ranking}

\input{src/06-application}
\input{src/07-conclusion}

\begin{acks}
This work was partially supported by grants from CNPQ, FAPEMIG, and FAPESP.
\end{acks}

\bibliographystyle{ACM-Reference-Format}
\balance
\bibliography{references}
\balance

\end{document}

%% file: src/00-abstract.tex
WhatsApp has introduced a novel avenue for smartphone users to engage with and disseminate news stories. The convenience of forming interest-based groups and seamlessly sharing content has rendered WhatsApp susceptible to the exploitation of misinformation campaigns. While the process of fact-checking remains a potent tool in identifying fabricated news, its efficacy falters in the face of the unprecedented deluge of information generated on the Internet today. In this work, we explore automatic ranking-based strategies to propose a ``fakeness score'' model as a means to help fact-checking agencies identify fake news stories shared through images on WhatsApp. Based on the results, we design a tool and integrate it into a real system that has been used extensively for monitoring content during the 2018 Brazilian general election. Our experimental evaluation shows that this tool can reduce by up to 40\% the amount of effort required to identify 80\% of the fake news in the data when compared to current mechanisms practiced by the fact-checking agencies for the selection of news stories to be checked.

%% file: src/01-introduction.tex
\section{Introduction}

The emergence of WhatsApp has introduced an unparalleled communication channel that connects more than two billion users around the world\footnote{\url{https://twitter.com/WhatsApp/status/1364714386078621703}}.
It has dramatically changed how people consume and share news stories in many places. A report from the Reuters Institute found that 
WhatsApp has become a primary network for discussing and sharing news in several countries~\cite{reuters2019report}. 
However, the large size of the platform and the easiness to forward messages and create chat groups have made WhatsApp prone to abuse by misinformation campaigns and dissemination of false rumors~\cite{melo2019can}. 
The frequent use of WhatsApp for socialization also favors this behavior. Indeed, Reuters also reported  that 76\% of WhatsApp users are members of groups and 58\% participate in groups with people they do not know.
This is evident in countries like Brazil, where WhatsApp has been pointed out as one of the main vectors of fake news spreading during elections \cite{nyt2018benevenuto} as well as a major platform to distributing political messages in bulk \cite{magenta2018}.



An effective way to detect fake news disseminated on social platforms is direct fact-checking, typically performed by expert journalists. A fact-checking task (i.e., the assessment of the truthfulness of a news story or claim \cite{vlachos2014fact}) verifies the correctness of the information by comparing it with one or more reliable sources \cite{myslinski2012fact}. 
 However, fact-checking is a time-consuming process since it commonly requires a detailed analysis to support the verdict~\cite{vlachos2014fact}. Thus, traditional fact-checking cannot keep up with the increasingly enormous volume of information that is now generated online~\cite{ciampaglia2015computational}. 
On the specific case of WhatsApp, there is an additional challenge: the established communication channels are decentralized and mostly private, by default. Thus, one cannot know what trends are discussed there and journalists usually need help even to identify popular pieces of media that require verification.  
In this scenario, automatic solutions for fake news detection could be used as an assistive tool for fact-checkers to identify content that is more likely to be fake or content that is worth checking, still leaving the final call to an expert at the endpoint of the process. 

One such tool is the WhatsApp Monitor~\cite{melo2019whatappMonitor}, 
 a web-based system that helps researchers and journalists by ranking content shared on WhatsApp public groups and displaying them in an organized way to help in the fact-checking task. It integrates a framework that \textit{i)} collects data from hundreds of public groups on WhatsApp, monitoring multiple types of media such as images, videos, audio, and textual messages from different countries, including Brazil, India, and Indonesia; \textit{ii)} matches identical pieces of information together and ranks them in terms of popularity every day; and, \textit{iii)} displays the ranked content on a web application where users can navigate through days and content.
This system proved to be very useful during the Brazilian 2018 elections, in which it was used as data source by more than a hundred journalists and  three fact-checking agencies~\cite{melo2019whatappMonitor}.
However, despite its extensive practical use, the system design is in essence quite simple, as it only displays a list of the most frequently shared content in the monitored groups over a time interval. Using only this information to choose which content should be fact-checked first may be misleading, as popularity of a news story in the  monitored groups may not be representative of its popularity elsewhere. Also, popularity and fake news are clearly distinct subjects:  even though fake news tend to spread more quickly than true stories~\cite{Vosoughi1146}, being more popular does not imply in being fake, necessarily. 

Therefore, in this work, we propose a new ranking mechanism that accounts for the potential occurrence of fake news within the data and implement it in WhatsApp Monitor, significantly reducing the number of content pieces journalists and fact-checkers have to go through before finding a fake story. Specifically, we explore automatic ranking-based strategies to propose a ``fakeness score'' model as a means to help fact-checking agencies identify fake news stories shared through images on WhatsApp. 

Our study relies on a dataset gathered from WhatsApp public groups during the 2018 Brazilian election, a period of great social mobilization during which WhatsApp was a primary means of communication and information spread in the country \cite{resendeWWW19}. Since this is, to the best of our knowledge, the first effort that explores ranking-based strategies to automatically detect fake news on WhatsApp, we consider hand-engineered features proven to be effect on other platforms. We compute almost 200 features, including novel features not previously studied, and investigate their discriminative power for identifying fake news. We emphasize that exploring these features and their discriminatory potential is also a strategy to better understand the unique characteristics of fake news disseminated on WhatsApp in the Brazilian political context. We then explore machine learning methods to estimate a fakeness score on news stories aiming at improving ranking results, which can support decisions regarding the selection of facts (or news) to be checked. Lastly, we deploy our approach in a real system, i.e., the WhatsApp Monitor\footnote{\url{http://www.monitor-de-whatsapp.dcc.ufmg.br/}}. 
Our experimental evaluation shows that this tool can reduce by up to 40\% the amount of effort required to identify 80\% of the fake news in the data when compared to current mechanisms practiced by the fact checking agencies for the selection of material to be checked (e.g., popularity ranking).   

The rest of the paper is organized as follows: Sections~\ref{sec:related} and~\ref{sec:definitions} present related work, important definitions, and problem statement. We introduce our data gathered from WhatsApp and from fact-checked news stories (i.e.\ labeled data) in Section \ref{sec:datasets},  and describe our implemented features for fake news detection in Section~\ref{sec:features}. Section~\ref{sec:evaluation} describes our experimental evaluation and main results. Section~\ref{sec:applications} discusses potential applications of our approach, while Section~\ref{sec:conclusion} concludes the paper and points some directions for future work.

%% file: src/02-background.tex
\section{Related Work}\label{sec:related}
There is a vast group of efforts that propose solutions to mitigate the fake news problem based on various artificial intelligence techniques such as classic supervised approaches~\cite{conroy2015automatic}, active learning~\cite{bhattacharjee2017active} and deep learning~\cite{wang2018eann}. Despite the emergence of many computational solutions to detect fake news, manual fact-checking done by expert journalists is still in large use as it remains as the only  completely reliable solution. Indeed, manual fact-checking is particularly important in platforms like WhatsApp, in which access to content is more limited.

WhatsApp has become a powerful instrument to influence people during political campaigns. The app has been reportedly abused by misinformation campaigns, especially in countries from South America, Africa, and Southeast Asia~\cite{reuters2019report}. To better understand this phenomenon, researchers have  collected~\cite{garimella2018whatapp} and analyzed WhatsApp data under various lenses, such as user behavior in a biased political scenario~\cite{bursztyn2019thousands} and collaborative production of information as means to engage users~\cite{lambton2019whatfutures}. 
There are also studies that aim at better comprehending the phenomenon of misinformation spread on WhatsApp, focusing on  content patterns and propagation dynamics~\cite{resendewebsci19,resendeWWW19}.
Particularly, Resende \textit{et al.}~\cite{resendewebsci19,resendeWWW19} analyzed the dissemination of misinformation in images and textual content, focusing on publicly accessible, politically-oriented groups. They collected messages shared during two major social events in Brazil and found a large presence of misinformation in the shared content by using labels provided by journalists. 
 Motivated by the massive spread of misinformation and rumors on the application, WhatsApp engineers have been deploying measures to mitigate this problem, such as reducing the maximum number of users to which a message can be forwarded at once. Yet, it has been shown that, although the current changes deployed by WhatsApp can indeed greatly delay the information spread, they are ineffective in blocking the spread of misinformation   through public groups when the content has a high viral nature~\cite{melo2019can}.

Along the lines of proposing countermeasures to misinformation spread, WhatsApp Monitor~\cite{melo2019whatappMonitor} provided a relevant platform to explore content shared in hundreds of public groups in Brazil, India, and Indonesia. This tool is able to group similar content by using different heuristics (e.g.,  perceptual hash~\cite{monga2006perceptual} for image content) on a large volume of data and rank it by popularity (i.e., number of shares) daily. This ranking was used by many journalists and agencies  during the 12 weeks leading up to the Brazilian 2018 presidential election, including Comprova\footnote{\url{https://projetocomprova.com.br/}}, a collaborative journalistic project led by First Draft focused on verifying questionable stories published on social media and WhatsApp.  
In this work, we use a dataset priorly collected from WhatsApp to explore new strategies for detecting fake news disseminated on this platform and thus help fact-checkers identify, at an early stage, content that is more likely to be fake. 


Although there are many efforts that study fake news detection \cite{gangireddy2020unsupervised, gaglani2020unsupervised, luvembe2023dual}, including some isolated initiatives in Brazil~\cite{monteiro2018contributions, reis2021supervised}, they are limited in terms of evaluating the viability of actual deployment of these solutions in a real-world system. Thus, different from previous efforts, we implement our proposed approach in a real system (i.e. WhatsApp Monitor) that, as aforementioned, has been used extensively for monitoring the Brazilian electoral process. Further, to the best of our knowledge, this is the first work that explores strategies with practical potential for detecting fake news spread on WhatsApp.



\section{Definitions and Proposed Framework}\label{sec:definitions}




We here consider the problem of extracting and leveraging information associated with content and propagation properties in WhatsApp to build a ranking tool that provides a sorted list where content most likely to be associated with a {\it fake story} appears at the top. We focus specifically on {\it image} content. From a technical standpoint, we address a ranking problem: given a set $S$ of news stories disseminated through images on WhatsApp, our goal is to generate a list $R$ in which news stories are sorted in descending order according to the likelihood of being fake. 
While addressing the aforementioned problem, we adopt the following definition of ``fake news": \textit{``A news article or message published and propagated through media, carrying false information regardless the means and motives behind it"}~\cite{sharma2019combating}.   

\begin{figure*}[t!]
	\centering
  		\includegraphics[width=0.7\textwidth]{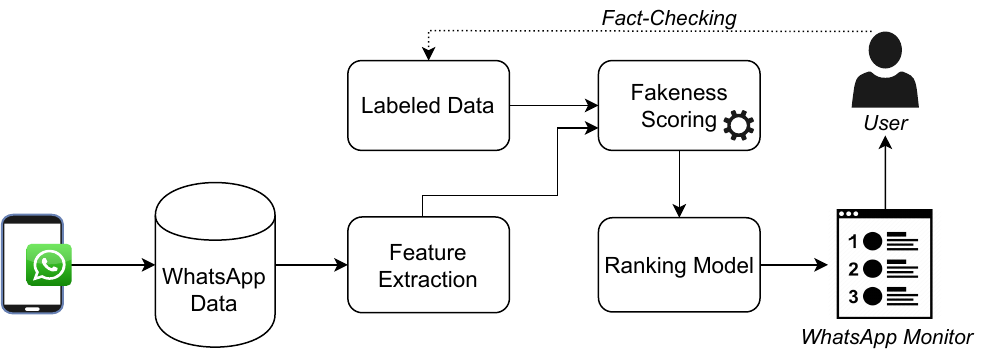}
	\caption{Framework for ranking WhatsApp content w.r.t.\ fakeness scores.}
	\label{fig:architecture}
\end{figure*}

The architecture of the proposed ranking tool is organized into the modules shown in Figure \ref{fig:architecture}. The core component is a model that, given a message as input, outputs a ``fakeness score".  To that end, this model explores almost 200 different features, some of which had been explored in the past and others are novel attributes proposed by us, extracted from the input message (Feature Extraction).  A labeled dataset,  composed of messages priorly checked as either fake or not (Labeled Data),  is required to learn the model to estimate the score (Fakeness Scoring). This model can then be used to compute scores for new messages  based on the extracted features, and rank them in decreasing order of score (Ranking Model). Lastly,  we make this ranking available in the WhatsApp Monitor for users, including potential fact-checkers, who can, in turn, eventually help us improve the model by providing more labeled data. In the following sections, we describe each of the components of this tool in detail. 


%% file: src/03-datasets.tex
\section{Datasets}\label{sec:datasets}

In order to evaluate the practical potential of our proposal, we need a large dataset of content collected from WhatsApp and also a large dataset of news stories labeled by fact-checkers that allows us to identify which content shared on WhatsApp is fake. In this section, we explain \textit{i)} how we gathered content from public politically-oriented WhatsApp groups in Brazil, and \textit{ii)} how we built labeled data from publicly available fact-checking websites. For the sake of designing and evaluating our strategy, {\it we focus on image content shared in WhatsApp}\footnote{Previous efforts showed that images are the most frequent type of media content, as well as an important source of fake news~\cite{resendeWWW19}.}, although it can be extended to other content types as well. 

\subsection{WhatsApp Data} 

WhatsApp is a very popular platform, but it remains barely explored. Due to the sheer volume of information shared in this messaging app and its closed nature (as it uses peer-to-peer encryption), getting relevant data from this network is a challenge by itself. It is difficult to have a good overview of this data or even to tell which items are popular. Hence, efforts towards helping organize and provide a view of the WhatsApp network for journalists and researchers are very valuable.

To achieve the goal of helping users navigate through the content and be informed about the relative fakeness of each image, we use a dataset built in~\cite{resendeWWW19},
which consists of messages containing images that were posted on WhatsApp during the 2018 Brazilian elections. This period is of particular interest to our study due to the great social mobilization and the reported large use of WhatsApp as a primary means for (mis-)information  dissemination in the country at the time, which drove a great effort from fact-checking agencies to publicize verified (checked) data. Thus, we were able to gather a reasonably large dataset of fake images, some of which were shared many times during the period, which allowed us to assess the effectivness of our  model.




The dataset consists of 4,524 distinct images disseminated on WhatsApp between August and October 2018. These images appeared in a total of 414 unique groups and were shared by 17,465 different users. In order to build it, the authors of~\cite{resendeWWW19} searched the Web for invite links to WhatsApp public groups, joined those related to the 2018 Brazilian elections and collected the shared messages. 
To make account of the popularity of each image, their methodology consists of identifying and counting duplicates of a same image by adopting the \textit{Perceptual Hashing (pHash)} algorithm~\cite{monga2006perceptual}. The pHash calculates a hash fingerprint for each image based on its visual features and then groups the images having the same hash values. In this way, they could track the dissemination and shares of a unique image through WhatsApp.

Our goal is to explore fake news present in the images in this dataset. 
To that end, all distinct images were labeled by using content previously fact-checked 
by specialized agencies in Brazil, as described next.  


\subsection{Labeled Data} 


A key step in automating fake news detection is obtaining a high quality dataset labeled by annotators with domain expertise. In~\cite{resendeWWW19}, the authors built a set of fake news that were
spread on WhatsApp, by obtaining images that were previously verified by specialized fact-checking agencies. More specifically, they submitted each image in the collected dataset to  the Google reverse image search and checked whether one of the main fact-checking domains in Brazil\footnote{\url{aosfatos.org}, \url{boatos.org}, \url{e-farsas.com}, \url{g1.globo.com/e-ou-nao-e/}, \url{piaui.folha.uol.com.br/lupa/}, \url{veja.abril.com.br/blog/me-engana-que-eu-posto/}} 
were returned in the results. If so, the fact-checking page was parsed and they automatically labeled the image as fake according to how the image was tagged by the fact-checking page. Messages containing images in which the truthfulness of the content could not be checked (i.e., were not returned in the search results)  were referred to as  \textit{unchecked messages}.

Here, we extend this process using Google Vision API\footnote{\url{https://cloud.google.com/vision}} to obtain the related domains in which each image has also appeared. For each result containing a fact-checking domain, we automatically label the corresponding image following the verdict of the agency. 
In order to ensure the quality of this data, we manually verified that all images labeled as fake by this methodology were found in fact-checking websites and matched those in our WhatsApp data. It is important to mention that our dataset stands out for its creation and the labeling process is already presented in more detail in another work of ours \cite{reis2020dataset}. In sum, our final dataset of images labeled as ``previously fact-checked'' contains 135 distinct images which correspond to fake news\footnote{The data is publicly available here: \url{http://doi.org/10.5281/zenodo.3779157}}. 
In the next section, we present the set of features extracted from this dataset and show that they can be useful to distinguish fake news from unchecked content. 
\vspace{0.15cm}
\noindent 
\textbf{\\Data Limitations.} 
\textit{i)} The dataset used comprises only publicly accessible WhatsApp groups. This is a small fraction of all WhatsApp network, yet this is the largest sample from WhatsApp  available, to our knowledge, for exploring fake news detection. 
\textit{ii)} Some of the images marked as unchecked may actually be fake due to one of two reasons. First, many fact-checking agencies do not post the image that has been disseminated, preventing us to match these fact-checks with our data. Second, the stories behind some of the unchecked messages may not have been verified (e.g.\ fact-checkers tend to favor the verification of most shared content).
\vspace{0.15cm}
\noindent 
\textbf{\\Ethical considerations.} 
 This work does not involve experiments with human subjects.  All sensitive information in the WhatsApp dataset (i.e., group names and phone numbers) has been previously anonymized to ensure the privacy of users. Also, the fact-checking data is publicly available, and only includes highly shared images. 

%% file: src/04-features.tex
\vspace{-0.05in}
\section{Feature Extraction}\label{sec:features}
Previous efforts showed that features for fake news detection can be divided into three main groups~\cite{reis2019supervised}: (i) features extracted from \textit{content} (e.g., textual and image properties); (ii) features extracted from  the \textit{source} (e.g., who is the publisher? trustworthiness); and (iii) features extracted from the \textit{environment}, which usually involve propagation dynamics within social platforms and the Web. Table~\ref{tab:overviewfeatures} summarizes the main features for fake news identification that we implemented in this work, divided by each group. In total, we have 181 features for fake news detection. 
In addition to features typically used for this task, we proposed features based on image properties, semantic structure, and specific attributes of publishers and propagation within/outside WhatsApp, that have not yet been explored for detecting fake news within this social platform. 
Next, we briefly describe the features for fake news detection included in each group and assess their relative ability to identify fake news disseminated on WhatsApp.

\begin{table*}[!t]
\centering
\caption{Feature overview.} 
\resizebox{\textwidth}{!}{%
\begin{tabular}{|l|l|p{11cm}|c|}
\hline
Extracted from... & Feature Set & General Description & Total \\ \hline
\multirow{6}{*}{Content} & Image properties & Number of faces in image, labels, colors, objects, etc. & 9$^{\ddag}$  \\ \cline{2-4} 
                                                      & Language structures (Syntax)      & Sentence-level features, indicators of text quality (e.g. readability metrics) & 31                         \\ \cline{2-4} 
& Lexical features                  & Character level and word-level features, including number of words, first-person pronouns, demonstrative pronouns, verbs, hashtags, all punctuations counts, etc. & 49                         \\ \cline{2-4} 
                                                       & Psycholinguistic cues             & Additional signals of persuasive language such as anger, sadness, etc and indicators of biased language                                                           & 38                         \\ \cline{2-4} 
                                                       & Semantic structure                & Labels, contextual informations, toxicity score                                                                                                                   & 8$^{\star\ddag}$                         \\ \cline{2-4} 
                                                       & Subjectivity cues                 & Subjectivity and sentiment scores                                                                                                                                 & 4                          \\ \hline \hline
\multirow{2}{*}{Source}                              & Publisher                         & The user who carried out the first news share, the groups where this news was disseminated                                                                        & 5$^{\star\ddag}$                         \\ \cline{2-4} 
& Bias                              & Political alignment measure (i.e. left, right, mainstream)                                                                                                      & 3                         \\ \hline \hline
\multirow{3}{*}{\begin{tabular}[l]{@{}l@{}}Environment \\ (Social Platforms \\ and Web)\end{tabular}} & Internal propagation (Engagement) & Number of shares, distinct users who posted the same message, and distinct groups in which the same message was posted                                            & 3$^{\ddag}$                          \\ \cline{2-4} 
& External propagation              & Information about the spread of the story outside whatsapp                                                                                                        & 5$^{\ddag}$                          \\ \cline{2-4} 
                                                       & Temporal Patterns                 & The rate at which shares are made internally for different time windows                                                                                           & 26                          \\ \hline
\multicolumn{3}{l}{The presence of $^{\star}$ indicates the existence of categorical features, and $^{\ddag}$ indicates features not previously used for fake news detection.}\\
\end{tabular}%
}
\label{tab:overviewfeatures}
\end{table*}

\subsection{Content Features}

Content features involve not only the news story but also its headline, associated images, and any message that was published along with it. Here, since we consider only news stories disseminated on WhatsApp through images
, we first use the Google Vision API to extract various pieces of information associated with them, e.g., \textbf{image properties} such as labels, colors, objects, and  the presence of faces. We also used as features measures provided by safe search detection, which detects explicit content (adult, spoof, medical, violence, and racy) and returns the likelihood that each is present. This approach is a simpler alternative to training a specialized network (e.g., CNN) to extract image features.

Additionally, we extract text from images using the optical character recognition (OCR) provided by the same API to compute features for fake news detection proposed by previous efforts such as \textbf{language structures} (e.g., sentence-level features including number of words and syllables per sentence, indicators of text quality from readability metrics), \textbf{lexical features} such as character and word-level signals, among others.

We also use the $2015$ version of the Linguistic Inquiry and Word Count (LIWC) \cite{tausczik2010psychological} to extract and analyze the distribution of \textbf{psycholinguistic cues}\footnote{LIWC is a psycholinguistic lexicon system that categorizes words into psychologically meaningful groups organized as a hierarchy of (sub-)categories, which form the set of LIWC attributes.} from the text included in and associated with an image. Since its conception, LIWC has been widely used for a number of different tasks, including discourse characterization in social media platforms and fake news detection~\cite{reis2019supervised}. We then extracted aspects of the \textbf{semantic structure} of a text including contextual information and indicators of toxicity\footnote{The Perspective API, available at \url{https://www.perspectiveapi.com} uses machine learning models to quantify the extent to which a text can be perceived as “toxic".}. Last, 
we also computed \textbf{subjectivity features}, namely subjectivity and sentiment scores.

\vspace{-0.05in}
\subsection{Source Features}

Features from source are related to the publisher of the news story, e.g., domain information extracted from news URLs. Here we consider as \textbf{publishers} the users and groups within WhatsApp to capture information about the potential fabricators of news stories. 

First, we consider the anonymized identifier of the user who shared a news story for the first time as a categorical feature. Similarly, we capture the first WhatsApp group in which it was posted. In a preliminary analysis, we found that only 10 out of more than 17K users were responsible for the first posting of 23\% of images fact-checked as fake, and 9 out of 414 groups concentrate nearly half (44\%) of the first time appearances of fake images. We conjecture that these statistics provide valuable information, capable of capturing any indication of a malicious and orchestrated action to intentionally spread fake news.

Previous efforts show a correlation between political polarization and the spread of fake news~\cite{ribeiro2017everything}. Thus, we infer the political \textbf{biases} of WhatsApp groups according to the following strategy: (1) we automatically parsed the group description (i.e., group name) to check if there was any information about its political bias. If so, we label the group as ``right'', ``left'' or ``mainstream''. 
For example, the group ``\# BOLSONARO PRESIDENTE'' was assigned ``right'' as political bias since Jair Bolsonaro, back then a presidential candidate, is a right-wing partisan. For cases where the description of the group did not provide any indication of its political alignment, (2) we manually inspect the group content, that is, the bias of messages posted in them, in order to infer the political bias of group. This strategy has been used in previous studies to quantify the biases of a given source~\cite{budak2016fair,ribeiro2018@icwsm}. For cases where content from both biases was shared across the same groups, we labeled it mainstream. In a preliminary analysis, we found that during the 2018 electoral period, right-wing groups were more active in the dissemination of content within WhatsApp. Furthermore, an image posted in those groups is more likely to be associated with a fake story than one posted in another group due to the imbalance between fake and unchecked content. 
This corroborates previous studies showing that right-wing groups are more effective in using the social media tool to spread news, disinformation and opinions~\cite{bursztyn2019thousands}.






\vspace{-0.05in}
\subsection{Environment Features} 
\label{sec:features_environment}

Some features can be extracted from the environment such as user engagement metrics and statistics from propagation dynamics. In this work, we computed \textbf{internal propagation} measures, i.e., the number of distinct users who posted the same news story through image on WhatsApp, the number of distinct groups in which the same news story was posted, and the total number of copies (shares) of the same news story across all analyzed groups both, for messages containing fake news and for messages with unchecked content.

Additionally, we recover \textbf{external propagation} measures, i.e., information about the dissemination of these WhatsApp on the Web. To accomplish this, we use the information about pages with matching images from the Google Vision API which returns information about websites that contain images identical to an image provided as input. From the set of websites/domains that published this image over the Web, we measured the volume of available, uncommon\footnote{To determine common links we used pre-defined suffixes: `.com', `.net', `.edu', `.org', `.mil', `.gov', `.br' from \url{https://www.domain.com/blog/2018/10/30/domain-name-types/}}, and secure links (i.e., https).

Last, to capture \textbf{temporal patterns} of each news story from sharing activity on WhatsApp, we compute the rate at which shares are made within intervals from first share time  (900, 1800, 2700, 3600, 7200, 14400, 28800, 57600, 86400, 172800, 259200, 345600, and 432000 seconds). Similarly to Twitter~\cite{Vosoughi1146}, we found that fake news have a much faster reach in WhatsApp and over the Web, suggesting a viral behavior within and outside the WhatsApp.

\subsection{Feature Importance}

In this section, we evaluate the relative power of each feature in discriminating fake news from the unchecked content by ranking them w.r.t.\ the Information Gain (IG)~\cite{MIR2011}.
Table~\ref{tab:feature_importance} lists the top-20 most discriminative attributes according to this measure. 

\begin{table}[]
\centering
\small
\caption{Feature importance.}
\begin{tabular}{lll}
   \hline
   & \textbf{Top features by InfoGain (IG)}                & (\%) \\ \hline
1  & count\_web\_dissemination\_urls (env:ext prop)           & 9.30  \\
2  & web\_dissem\_accessible\_links (env:ext prop)            & 5.10  \\
3  & web\_dissem\_foreign\_uncommon\_domains (env:ext prop)   & 4.30  \\
4  & acc\_259200 (env:int prop)                               & 3.80  \\
5  & count\_groups (env:int prop)                             & 2.90  \\
6  & sentence\_info\_syll\_per\_word (content:synt)           & 2.80  \\
7  & Dic (content:lexic)                                       & 2.10  \\
8  & Bridge (content:semantic)                                    & 2.00  \\
9  & ingest (content:psych)                                    & 1.90  \\
10 & count\_low\_word (content:lexic)                          & 1.80  \\
11 & toxicity (content:semantic)                                  & 1.80  \\
12 & Quote (content:lexic)                                     & 1.70  \\
13 & img\_faces (img\_has\_faces) (env:image prop)              & 1.70  \\
14 & img\_count\_labels (content:image prop)                        & 1.60  \\
15 & Sixltr (content:lexic)                                    & 1.60  \\
16 & img\_count\_objects (content:image prop)                       & 1.60  \\
17 & political\_bias\_right (source:bias)                    & 1.60  \\
18 & anger (content:psych)                                     & 1.50  \\
19 & number (content:lexic)                                    & 1.40  \\
20 & cogmech (content:psych)                                   & 1.40  
 \\ \hline
\end{tabular}
\label{tab:feature_importance}
\end{table}

Note that the 20 most discriminative features are distributed among the three categories, i.e., content, source, and environment, underlining the need to use all of them. Moreover, we observe a trend: the content features are the majority, followed by environment and then source-related ones. Although the number of features extracted from content is larger, this highlights how important they are for detecting fake news, especially those related to semantic aspects. For instance, there are several news items that are labeled as fake simply because they present information, in some cases even true, but out of context. Moreover, considering IG weights, we can observe that propagation related attributes appear among the top 5 of all features, confirming that they are indeed useful for fake news detection purposes.  These results are consistent with those shown in Section~\ref{sec:features_environment}.

%% file: src/05-ranking.tex
\section{Experimental Evaluation}\label{sec:evaluation}

In this section, we describe our ranking-based strategies and details of our experimental setup. At the end, we present and discuss the main results.
\subsection{Ranking-Based Strategies}



Aiming to measure the potential of the proposed features, we analyzed three different and widely used machine learning models in our experiments: Support Vector Machine (SVM)~\cite{joachims1998text}, Multilayer Perceptron (MLP)~\cite{lecun_nature2015} and XGBoost (XGB)~\cite{chen2016xgboost}. These models learn from training data how to assign a fakeness score to a news story disseminated through images on WhatsApp. Each instance $i$ of the training dataset is composed by a feature vector $X_i$, containing the values of the 181 features described in Table~\ref{tab:overviewfeatures}, and a label $y_i$ indicating whether $i$ is fake ($y_i$=$1$) or unchecked ($y_i$=$0$). Given an unseen news message $j$, the output is an estimate of the probability of $j$ being fake, which in turn is used to produce a ranking of news stories. 


Regarding the specific characteristics of the learning approaches, SVM constructs a hyperplane or set of hyperplanes in a high-dimensional space, which separates training examples with different labels. 
A Multilayer Perceptron (MLP) is a class of feedforward artificial neural network (ANN) in which the nodes (neurons) are disposed in multiple layers: an input layer, an output layer and in between $L$ hidden layers. 
Finally, XGBoost (XGB) is short for eXtreme Gradient Boosting. It consists of an ensemble of ``weak models'', typically decision trees, which are combined into a single, strong model, using a boosting technique. 


\subsection{Evaluation Metrics}

Since our goal is to build good rankings, we assess the effectiveness of our ranking-based strategies using (appropriate) metrics commonly adopted in Information Retrieval and ranking tasks: Precision, Recall, and Normalized Discounted Cumulative Gain (NDCG)~\cite{MIR2011} in the top $k$ positions of the ranking (Precision@k, Recall@k, and NDCG@k, respectively) for different values of $k$. In the fact-checking domain, $k$ represents the number of news that the fact-checker specialist can afford to inspect.

Let $S$ be the set of all news that we desire to check, $F \subseteq S$ be the set of fake news among them, $R$ be the produced ranking, and $R^k$ be the top-k positions of $R$. Precision@k is the fraction of fake news detected in the first $k$ positions of the provided rank, and  Recall@k is the fraction of all existing fake news in  $S$ that were indeed retrieved among the top-$k$ positions of the ranking, that is:

\begin{equation} \label{eq:precision}
\mathit{Precision}@k(R, F) = \frac{\mid R^k \cap F \mid }{k}
\end{equation}

\begin{equation} \label{eq:recall}
\mathit{Recall}@k(R, F) = \frac{\mid  R^k \cap F \mid }{\mid F \mid }
\end{equation}

 
The NDCG@k metric, in turn, emphasizes results in the top positions of the ranking. Let $\mathit{DCG@k}$ be the discounted cumulative gain in the first $k$ positions of the ranking, defined below in the equation on the left, where $f(i)$ is equal to 1 if the  $i$-th news message in $R$ is fake (i.e.\ it is in $F$), and 0 otherwise. The un-normalized and normalized discounted cumulative gain in the first $k$ recommendations are respectively as:
%
\begin{equation} \label{eq:dcg}
 \mathit{DCG@k(R, F)} = \sum_{i=1}^k \frac{f(i)}{\log_2(i+1)}
\end{equation}

\begin{equation} \label{eq:ndcg}
 \mathit{NDCG@k(R, F)} = \frac{DCG@k(R, F)}{IdealDCG@k}
\end{equation}
%
%
%
\\
where $\mathit{IdealDCG}$ is the value obtained for $\mathit{DCG@k}$ when there are only fake news at the top-k (or fewer) positions.


\subsection{Experimental Setup}


We evaluate the fake news ranking methods using 5-fold cross-validation experiments. That is, the set of news stories is split into five equal-sized portions. Three portions are used as training set, to ``learn'' the models. One portion is used as validation set, for parameter tuning, and the last portion is used for testing. The portions are rotated such that 5 configurations of training, validation and test sets are tested.

We opted for a stratified folding methodology\footnote{We also exploited non-stratified random samples. Despite some variations in the absolute values of our results, the same conclusions are maintained for both configurations.}, ensuring that all samples have approximately the same fake/unchecked ratio, since the number of news labeled as ``fake'' is small (3\% of our dataset). In order to obtain multiple lists of news to rank, for each portion in each fold, we generated 50 bootstrap samples with replacement (also stratified), containing 200 news each. Thus, reported results are averages over 50 samples $\times$ 5 folds = 250 executions.


For each learning technique, we scaled the feature values using z-score normalization, and performed grid-search in the validation set in order to find the best parameter values. We chose the parameter values that led to the best NCDG@10, presented in Table \ref{tab:param}. However the results are similar using other evaluation metrics and ranking cutoffs.  


\begin{table}[]
\centering
\footnotesize
\caption{Parameterization of learning techniques.}
\begin{tabular}{llll}

\hline
Method     &  Parameter              &   Tested values & Choice \\
\hline

MLP        &  Number of neurons per layer & {20, 100, 200}               & 100 \\
           &  Number of hidden layers                    & {1, 2, 3, 5, 10}                   & 1   \\
           &  Learning rate            & {0.001, 0.01, 0.1, 0.3, 1}    & 0.3 \\
\hline
   
SVM        &  kernel        & {Linear, RBF}           &    RBF \\
           &  Regularization parameter ($c$)                    & {0.001, 0.01, 0.1, 1}   &    0.1 \\
\hline

XGBoost    &  Max. depth of produc. decis. trees      & {6, 10, 15}           & 10    \\
           &  Learning rate            & {0.001, 0.01, 0.1, 1} & 0.001 \\

\hline

\end{tabular}
\label{tab:param}
\end{table}


\subsection{Results}

Table~\ref{tab:results} shows average Precision@k, Recall@k and NDCG@k results (for $k$=5, 10, 50 and 100) for each method, namely, a baseline in which news are ranked by their number of shares ($\#\mathit{Shares_{Rank}}$), and the machine learning based techniques MLP, SVM and XGBoost. 95\% confidence intervals are also presented, and best results (with corresponding statistical ties according to a 2-sided t-test) are shown in bold. 

Our first observation is that, as expected, ranking news stories by the number of shares (although it is a good evidence of the impact the fake news may cause) is not ideal for finding fake news, performing poorly in comparison to the other strategies, w.r.t.\ all evaluation metrics. The best overall ranking strategy (XGBoost) outperforms this popularity-based baseline with gains of up to 64\%, 64\% and 77\% in Precision@10, Recall@10 and NDCG@10, respectively. 

For small values of $k$ (e.g., $k$=$5$ and $k$=$10$), 
both XGBoost and SVM produce the best results, being statistically tied. However, for $k=50$ and 100, XGBoost significantly outperforms the second best method (SVM) with gains of up to 27\%, 27\% and 23\% in precision, recall and NDCG, respectively.

\begin{table*}[]
\centering
\caption{Average experimental results and 95\% confidence intervals. Best results (and statistical ties) in \texttt{bold}.}
\begin{tabular}{lllll}

\hline
                   &     Precision@5   &    Precision@10    &    Precision@50    &    Precision@100  \\
\hline

$\mathit{\#Shares_{Rank}}$  & 0.092 $\pm$ 0.020          & 0.105 $\pm$ 0.014 &          0.064 $\pm$ 0.004 &           0.045 $\pm$ 0.002 \\ 
MLP                         & 0.153 $\pm$ 0.023          & 0.129 $\pm$ 0.014 &          0.061 $\pm$ 0.004 &           0.045 $\pm$ 0.002 \\
SVM                & \textbf{0.211} $\pm$ 0.029 & \textbf{0.166} $\pm$ 0.017 &          0.067 $\pm$ 0.004 &           0.045 $\pm$ 0.001 \\
XGBoost            & \textbf{0.237} $\pm$ 0.025 & \textbf{0.173} $\pm$ 0.016 & \textbf{0.085} $\pm$ 0.003 &  \textbf{0.053} $\pm$ 0.001 \\

\hline
                   &      Recall@5    &       Recall@10   &    Recall@50    &      Recall@100       \\
\hline

$\mathit{\#Shares_{Rank}}$ & 0.077 $\pm$ 0.017 & 0.176 $\pm$ 0.023 & 0.530 $\pm$ 0.033 & 0.744 $\pm$ 0.031 \\
MLP                 & 0.128 $\pm$ 0.019 & 0.216 $\pm$ 0.023 & 0.504 $\pm$ 0.031 & 0.744 $\pm$ 0.027 \\
SVM                 & \textbf{0.176} $\pm$ 0.024 & \textbf{0.277} $\pm$ 0.028 & 0.556 $\pm$ 0.031 & 0.748 $\pm$ 0.024 \\
XGBoost             & \textbf{0.198} $\pm$ 0.021 & \textbf{0.288} $\pm$ 0.027 & \textbf{0.706} $\pm$ 0.027 & \textbf{0.891} $\pm$ 0.020 \\

\hline
                   &      NDCG@5      &      NDCG@10     &      NDCG@50     &      NDCG@100      \\
\hline

$\mathit{\#Shares_{Rank}}$  & 0.110 $\pm$ 0.025 & 0.113 $\pm$ 0.017 & 0.076 $\pm$ 0.006 & 0.057 $\pm$ 0.003 \\
MLP                 & 0.152 $\pm$ 0.024 & 0.136 $\pm$ 0.016 & 0.077 $\pm$ 0.006 & 0.059 $\pm$ 0.003 \\
SVM                 & \textbf{0.213} $\pm$ 0.031 & \textbf{0.181} $\pm$ 0.021 & 0.092 $\pm$ 0.007 & 0.066 $\pm$ 0.004 \\ 
XGBoost             & \textbf{0.250} $\pm$ 0.029 & \textbf{0.201} $\pm$ 0.020 & \textbf{0.113} $\pm$ 0.006 & \textbf{0.078} $\pm$ 0.004 \\

\hline

\end{tabular}
\label{tab:results}
\end{table*}

Thus, our results show that our features can be effectively used for learning a good fakeness function, being XGBoost the best  method considering all evaluation metrics. In the next section, we analyze our results under the lens of fact-checkers, which leverage our methods to detect fake news earlier.


%% file: src/06-application.tex
\section{Potential Applications}\label{sec:applications}

We now discuss potential applications of the ``fakeness score'' model proposed in this work, focusing on a analysis of costs to fack-checkers (Section~\ref{sec:cost}) and its deployment in a real application (Section~\ref{sec:monitor}).

\subsection{Fact-Checking (Cost Analysis)} \label{sec:cost}

One of the most important applications of the tool we propose here is the support to the fact-checking process. Ranking images in the WhatsApp Monitor according to a fakeness score can be used to help an expert fact-checker assign priorities in a more informed way.

We now revisit the previous results from the fact-checker perspective. Specifically,  for each strategy we evaluate (i) the effort required to recover a given fraction of the fake news and, conversely, (ii) the fraction of fake news recovered upon fact-checking a given fraction of the images. We assume that the fact-checker  follows the ranking returned by each strategy and that the cost of checking any image is fixed and the same.

\begin{figure}[t!]
	\centering
		\includegraphics[width=0.45\textwidth]{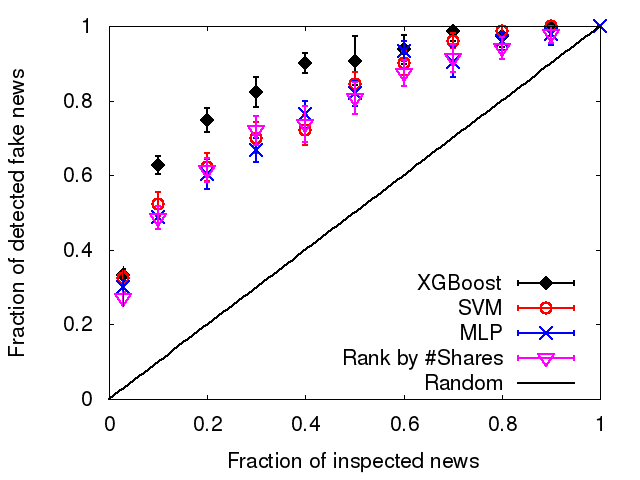}
	\caption{Cost analysis.}
	\label{fig:costanalysis}
\end{figure}

Figure~\ref{fig:costanalysis} shows, for each strategy, the fraction of news inspected (x-axis) and the average fraction of detected fake news with confidence intervals (y-axis). We observe that to recover 80\% of the fake news, a ranking created with XGBoost would require a fact-checker to check approximately 30\% of the news, while other strategies would require around 50\%. Interestingly, by checking the top 40\% entries ranked by XGBoost, the fact-checker would recover roughly 90\% of all fake news in the dataset. These results show that the use of ML-based methods for estimating a fakeness score can significantly reduce the efforts required to identify fake news, potentially allowing them to be found at an early stage.


\subsection{WhatsApp Monitor} \label{sec:monitor}

Finally, 
we implemented and deployed the fake scoring and ranking model in a real application for exhibiting WhatsApp data, extending the system proposed in \cite{melo2019whatappMonitor}. Their work provides the first version of an online system that exposes the daily trends shared on WhatsApp public groups for a particular country (e.g., Brazil) or domain (e.g., politics and news).
It is available to journalists and researchers seeking to analyze WhatsApp data. With our proposed extension, which allows users to rank images based on their likelihood to be fake,  we hope to assist fact-checking agencies to fight misinformation.    

\begin{figure}[t!]
	\centering
	\subfigure[Navigation search]{\includegraphics[width=0.78\linewidth]{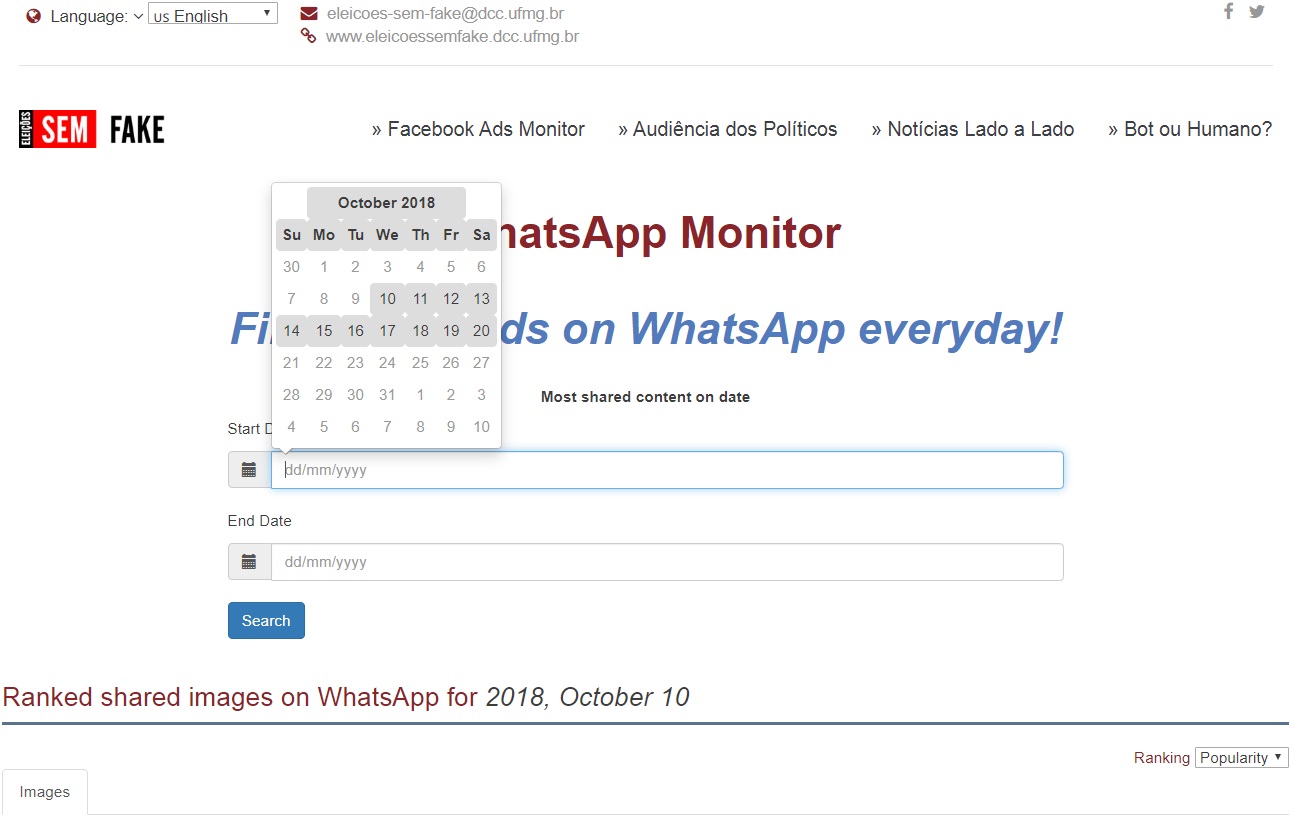}\label{fig:monitor1}}

	\subfigure[Image details]{\includegraphics[width=0.78\linewidth]{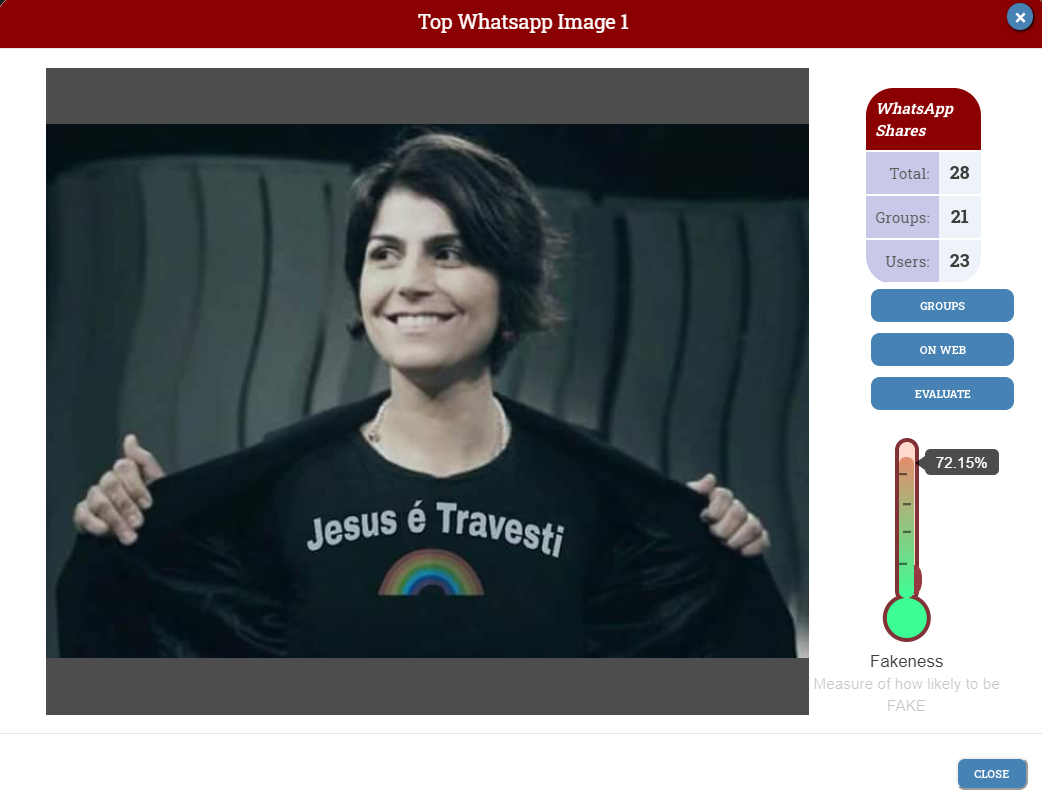}
	\label{fig:monitor2}}
	\caption{Screenshots of WhatsApp Monitor interface.} 
	 \label{fig:screenshotsmonitor}
\end{figure}

Figure~\ref{fig:screenshotsmonitor} presents some screenshots of the WhatsApp Monitor interface
\footnote{A demo of the tool can be accessed in \url{http://150.164.1.202/test_monitor/demo/}.}. 
After accessing the tool and logging in, users are taken to a dashboard where they can navigate between dates and observe the most shared multimedia content in our monitored groups for a given date, as shown in Figure~\ref{fig:monitor1}. With the new implemented functionality, users can choose between different methods for ranking the content: by popularity, which sorts images by the total number of shares;  or by ``fakeness'', which is based on the probability that the image is fake, as estimated by our XGBoost model. The system also has two other ranking methods based on the number of different groups or different users who shared each image.
This should belp journalists identify the pieces of content that are more  worth checking each day. Also, using this new tool, any ranking model can be used as an ``off-the-shelf'' strategy. That means, different ranking models can be trained and aggregated to the system (e.g., a model trained with pornography or specific fake news related to health), and it is up to the users to choose how to rank the content based on their preferences.

By clicking on ``Details'' of an image (Figure~\ref{fig:monitor2}), it shows the numbers of shares, users, and groups in which the image appears, to help them identify some context associated with the content, and a visual representation of the fakeness score as a thermometer, along with the probability  assigned by the trained model. 
To ensure the privacy of users, we do not  disclose any Personally Identifiable Information (PII) such as phone numbers. We use only ids in order to measure aggregate spreading statistics. Also, to avoid any misuse of the system, we limit user access through a login account. Since we only use publicly available WhatsApp groups we joined, our data collection does not violate WhatsApp terms of service.

\begin{figure}[t!]
	\centering
	\subfigure[Popularity ranking]{\includegraphics[width=0.75\linewidth]{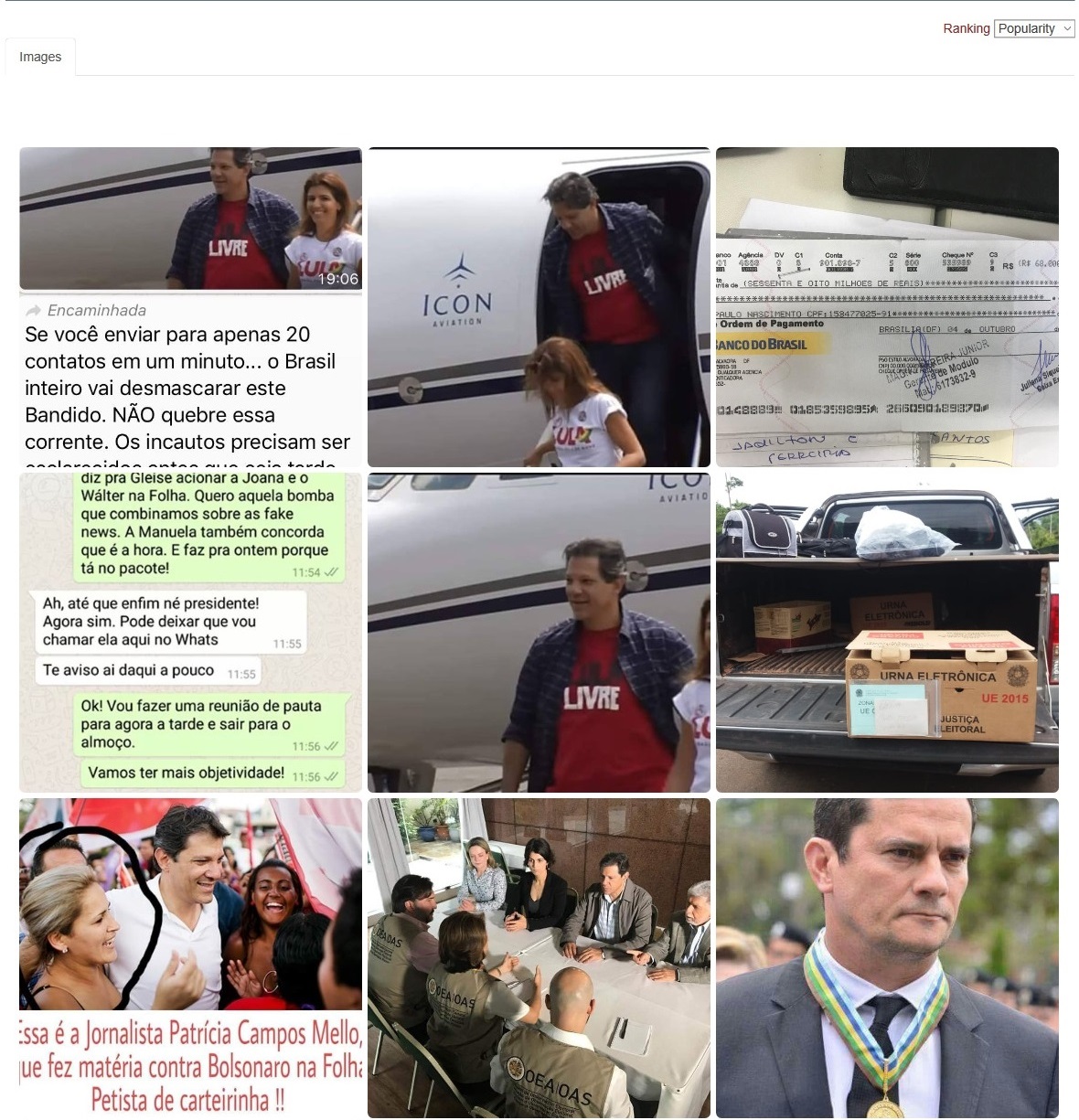}\label{fig:monitorshares}}
	\subfigure[Fakeness ranking]{\includegraphics[width=0.75\linewidth]{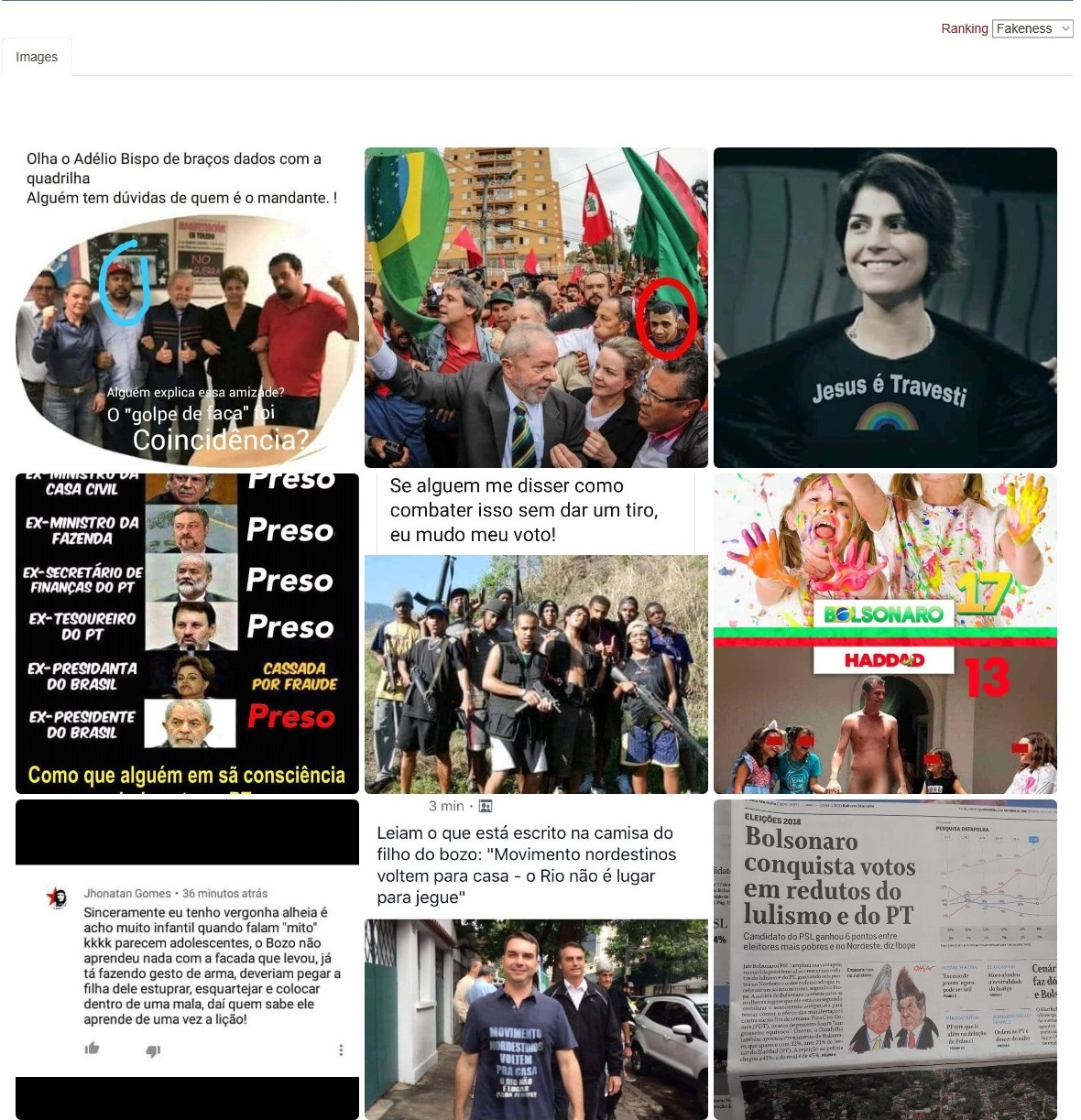}
	\label{fig:monitorfake}}
	\caption{Screenshots of the WhatsApp Monitor comparing the different ranking strategies for the same period.} 
	 \label{fig:rankingmonitor}
\end{figure}

Figure~\ref{fig:rankingmonitor} illustrates how images are displayed on the system, contrasting the results of the rankings of popularity and  fakeness. Note that the two criteria generate very distinct views of data. The ranking by fakeness~(Figure \ref{fig:monitorfake}) presents fake news in all top items in this example. The popularity ranking~(Figure \ref{fig:monitorshares}), in turn, shows the most shared images, but only 4 of them were priorly  verified to be fake. 
Note that, although the images displayed in the figure and the data used to train our model come from the same source, we picked a different date range than the one used for training the model to ensure a fair comparison.


%% file: src/07-conclusion.tex
\vspace{-0.05in}
\section{Conclusion}\label{sec:conclusion}

In this work, we consider the problem of fake news dissemination in WhatsApp images by investigating how to design and integrate a tool to the WhatsApp Monitor system that allows users to rank those images according to an estimated ``fakeness score''. To achieve this goal, we match data collected from the public groups on WhatsApp to news stories verified by fact-checking websites and evaluate the effectiveness of different ranking models or strategies. In this process, we implement 181 features that extract content, source and environment information to fit those models. Our experimental evaluation shows that our approach improves upon current, popularity-based mechanisms adopted by the checking agencies by up to 64\% in Precision and Recall. Moreover, the choice of classifier significantly impacts performance: XGBoost outperforms the second-best model by up to 27\%.

We discussed potential applications of this tool to fact-checking. In our experiments, the proposed tool reduced by up to 40\% the amount of effort required to identify 80\% of the fake news, hence significantly contributing to the fact-checking process. Moreover, we validate our approach by integrating the fakeness score model to a real system extensively used by Brazilian fact-checking agencies. As future work, we intend to conduct A/B tests to evaluate the effectiveness of the rankings generated by our approach in practice.

Finally, given the sheer volume and heterogeneity of data from WhatsApp w.r.t.\ content, source, etc, we plan to investigate whether it is possible to generate good rankings from reduced subsets of features in order to speed up the feature extraction process or even to improve results by reducing noise. In particular, it is useful to exclude source features for new users/groups. Moreover, we intend to explore more strategies to enhance the interpretability of our models \cite{reis2019explainable}.